\documentclass{article}

\usepackage{amsmath, amsthm, amsfonts}

\theoremstyle{definition}
\newtheorem{thm}{Theorem}[section]
\theoremstyle{theorem}

\theoremstyle{definition}
\newtheorem{defn}[thm]{Definition}

\theoremstyle{remark}

\usepackage{graphicx}
\usepackage{subfigure}          

\usepackage{placeins}

\usepackage{tabularx}

\usepackage{lmodern,textcomp}

\usepackage{algorithm}
\usepackage{algpseudocode}

\usepackage{enumitem}
\newlist{Aufz}{enumerate}{10}
\setlist[Aufz]{label*=\arabic*.}

\usepackage{xspace} 
\newcommand\largeparbreak{\par\bigskip}

\newcommand{\R}{\mathbb{R}\xspace}
\newcommand{\N}{\mathbb{N}\xspace}
\newcommand{\C}{\mathbb{C}\xspace}
\newcommand{\D}{\mathbb{D}\xspace}
\renewcommand{\H}{\mathbb{H}\xspace}

\newcommand{\cO}{\mathcal{O}\xspace}

\newcommand{\bx}{\textbf{x}\xspace}

\newcommand{\by}{\textbf{y}\xspace}

\newcommand{\hbx}{\hat{\textbf{x}}\xspace}

\newcommand{\be}{\textbf{e}\xspace}

\newcommand{\bO}{\textbf{0}\xspace}

\newcommand{\bv}{\textbf{v}\xspace}

\renewcommand{\t}{^{\textsf{T}}\xspace}

\newcommand{\comp}{\underline{i}\xspace}

\title{Review of theory and implementation of hyper-dual numbers for first and second order automatic differentiation}

\author{Martin Neuenhofen}

\date{\today}

\begin{document}

\maketitle

\begin{abstract}
	In this review we present hyper-dual numbers as a tool for the automatic differentiation of computer programs via operator overloading.
	
	We start with a motivational introduction into the ideas of algorithmic differentiation. Then we illuminate the concepts behind operator overloading and dual numbers.
	
	Afterwards, we present hyper-dual numbers (and vectors) as an extension of dual numbers for the computation of the Jacobian and the Hessian matrices of a computer program. We review a mathematical theorem that proves the correctness of the derivative information that is obtained from hyper-dual numbers.
	
	Finally, we refer to a freely available implementation of a hyper-dual number class in Matlab. We explain an interface that can be called with a function as argument such that the Jacobian and Hessian of this function are returned.
\end{abstract}


\section{Introduction}

\paragraph{Derivatives of computer programs}
Functions in computer programs can be often understood as implementions of mathematical functions. In this text we consider a function of the format
\begin{align*}
f \ : \ \C^n \rightarrow \C\,,
\end{align*}
where $n \in \N$ and $f$ is two times differentiable in a point $\bx \in \C^n$. There are many applications --- e.g. numerical optimization, sensitivity and stability analysis, risk analysis, numerical integration, cf. \cite{Naumann:Art} --- where 
first and second derivatives of such function $f$ are sought.

We introduce some notation on the derivatives. We write
\begin{align*}
	\nabla f(\bx)\t := \begin{bmatrix}
	\frac{\partial f}{\partial x_1}(\bx)& \dots &\frac{\partial f}{\partial x_n}(\bx)
	\end{bmatrix} \in \C^{1 \times n}
\end{align*}
for the \textit{Jacobian} of $f$ in $\bx$, and 
\begin{align*}
\nabla^2 f(\bx)\t := \begin{bmatrix}
\frac{\partial^2 f}{{\partial x_1}^2}(\bx) & \dots & \frac{\partial^2 f}{\partial x_1 \,\partial x_n}(\bx)\\
\vdots & \ddots & \vdots \\
\frac{\partial^2 f}{{\partial x_1}\,\partial x_n}(\bx) & \dots & \frac{\partial^2 f}{{\partial x_n}^2}(\bx)
\end{bmatrix} \in \C^{n \times n}
\end{align*}
for the \textit{Hessian} of $f$ in $\bx$.

\paragraph{Automatic differentiation}
Consider a function \texttt{f} within a computer program that implements the mathematical function $f$. Consider that $\nabla f(\bx)\t$ and/or $\nabla^2 f(\bx)$ shall be computed for a given $\bx \in \C^n$. Then there are two naive ways for obtaining this result:
\begin{enumerate}
	\item \texttt{f} is called in finite differences \cite{Fornberg:FD}.
	\item A programmer implements functions by hand that compute the Jacobian and Hessian of $f$.
\end{enumerate}
The first approach is undesirable since sensitive to rounding errors, cf. \cite{Fornberg:FD,Squire:1998:UCV:278129.278138}. Also the second approach is undesirable because it is error-prone and impractical to maintain when the implementation of $f$ is a subject of frequent changes.

Fortunately, there is a third approach called \text{automatic differentiation} \cite{GriewankAutoDiff}. Automatic differentiation aims at using an algorithmic approach that automatically generates and evaluates an analytic expression of the Jacobian and/or Hessian of $f$ at a given point $\bx$ although it is only given the implementation \texttt{f} (and $\bx$). To achieve this, algorithmic differentiation makes use of a wide range of mathematical approaches and programming paradigms. For a review we refer to \cite{Naumann:Art,GriewankAutoDiff}.

\paragraph{Dual numbers and operator overloading}
One particular approach in computational differentiation is the use of \textit{dual numbers} and \textit{operator overloading}. 

The idea behind dual numbers is that for each computed value within the code of \texttt{f} there is an auxiliary number that provides the derivative of this value. As an example to this, consider the following code, that implements a univariate function.
\begin{algorithmic}
	\Procedure{}{}{\texttt{foo}}({$x$})
	\State $u := \sqrt{x}$
	\State $v := x + u$
	\State $y := v/u$
	\State \Return $y$
	\EndProcedure
\end{algorithmic}
A dual number is a tuple of two numbers. We use the notation $\hat{x} = \langle x , \delta x \rangle \in \D$, where $\hat{x}$ is a dual number, consisting of $x,\delta x \in \C$, and $\D$ is the set of dual numbers. Consider now the following program
\begin{algorithmic}
	\Procedure{}{}{\texttt{bar}}({$\langle x,\delta x\rangle$})
	\State $u := \sqrt{x}$,\quad $\delta u := 0.5 \cdot x^{-0.5} \cdot \delta x$
	\State $v := x + u$,\quad $\delta v := \delta x + \delta u$
	\State $y := \hat{v}/\hat{u}$,\quad $\delta y := (v \cdot \delta u - u \cdot \delta v)/u^2$
	\State \Return $\langle y,\delta y\rangle$
	\EndProcedure
\end{algorithmic}
which is in fact identical to the former one, but has additional \textit{sensitivity equations}. Sensitivity equations mean that $\delta y$ will satisfy the following relation.
\begin{align*}
\delta y &= \nabla\texttt{foo}(x)\t \cdot \delta x
\end{align*}
So when $\delta x$ denotes the magnitude and sign of a perturbation in $x$ then $\delta y$ is the effect of this perturbation on $y$. The same holds for all intermediate results: $\delta u$ is the effect on $u$, $\delta v$ on $v$, etc. In the following we enlighten how the function \texttt{bar} can be evaluated even though only the function \texttt{foo} has been implemented.
\largeparbreak

In the following we introduce the concept of operator overloading. This concept is related to the paradigm of object-oriented programming \cite{Abadi:1996:TO:547964}. When implementing a new class, e.g. in \textsc{Matlab}, then all basic operators (such as $+$, $-$, $\times$, $/$,...) and basic functions (such as $\sin$, $\exp$, $\log$, $\sqrt{}$,...) must be defined from scratch for objects of this new class. This is because otherwise it is impossible for the computer to know how the programmer intended the class objects to be computed with each other. This process of redefining a function / operator (but for arguments of another class) is called \textit{function / operator overloading}, respectively.

When implementing a dual number class, it is suitable to overload the basic functions/operators $\sqrt{}$, $+$, $/$ in the following way:
\begin{align*}
	\sqrt{\langle a , \delta a\rangle\,} =: &
	\langle \sqrt{a}, 0.5 \cdot a^{-0.5} \cdot \delta a \rangle \\
	\langle a , \delta a\rangle + \langle b,\delta b\rangle =: &
	\langle a+b,\delta a + \delta b \rangle \\
	\langle a , \delta a\rangle / \langle b,\delta b\rangle =: &
	\langle a/b,(a \cdot \delta b - b \cdot \delta a)/b^2 \rangle
\end{align*}
Now, let us investigate a call of \texttt{foo} with a dual number $\hat x = \langle x , \delta x \rangle$. Since the arguments on the right-hand sides of the $:=$-operators are all dual, so will be the left-hand sides, respectively. Due to the above operator overloading, the call yields the following computations:
\begin{algorithmic}
	\Procedure{}{}{\texttt{foo}}({$\langle x,\delta x\rangle$})
	\State $\langle u,\delta u \rangle := \langle \sqrt{x} , 0.5 \cdot x^{-0.5} \cdot \delta x\rangle$
	\State $\langle v, \delta v\rangle := \langle x+u,\delta x + \delta u \rangle$
	\State $\langle y,\delta y \rangle := \langle v/u , (v \cdot \delta u - u \cdot \delta v)/u^2 \rangle$
	\State \Return $\langle y , \delta y \rangle$
	\EndProcedure
\end{algorithmic}
Comparing the expressions for $\delta u$, $\delta v$, $\delta y$, we find that they are identical to those in \texttt{bar}. So when calling \texttt{foo} with the dual number $\langle x, 1 \rangle$ input then the Jacobian of \text{foo} is readily available from the dual output $\langle \texttt{foo}(x), \nabla \texttt{foo}(x)\t \rangle$. This shall serve as a demonstration how dual numbers and operator overloading can be used for the computational differentiation of the function \texttt{foo}.

In practical applications we are interested in using dual numbers for the computational differentiation of a more general and complex (in terms of the number of lines of code) function \texttt{f}. This requires that all functions and operators that are called within \texttt{f} have been overloaded accordingly for dual numbers.

Yet we did not discuss how multivariate functions can be differentiated and how second derivatives can be computed.

\paragraph{The operator overloading rule}
At this point we briefly summarize the essentials of dual numbers.

A dual number is a tuple of numbers $\langle x,\delta x\rangle$. The left number $x$ stores the original value within the function evaluation, whereas $\delta x$ stores the sensitivity of $x$ with respect to the sensitivity of an input argument of the function call.

By means of operator overloading it is possible to incorporate sensitivity equations, so that these are evaluated whenever the function is called with a dual number input argument.

It remains the question of a specific rule that states how a basic function / operator must be overloaded. We answer this for a uni-variate function / operator. Suppose $g$ is such a basic function / operator, e.g. the $\sin$-function. Then $g$ must be overloaded for dual numbers, such that the following holds.
\begin{align*}
	\langle y,\delta y\rangle &:= g(\langle x,\delta x\rangle )\\
	y & := g(x)\\
	\delta y &:= \frac{\mathrm{d} g}{\mathrm{d}x}(x) \cdot \delta x
\end{align*}
For the example of the $\sin$-function, we obtain:
\begin{align*}
	\sin( \langle x , \delta x \rangle ) =: \langle \sin(x) , \cos(x) \cdot \delta x \rangle
\end{align*}
For a basic function / operator of two arguments one must use the following rule instead:
\begin{align*}
	\langle y,\delta y\rangle &:= g( \langle u,\delta u\rangle,\langle v,\delta v \rangle )\\
	y & := g(u,v)\\
	\delta y &:= \frac{\partial g}{\partial u}(u,v) \cdot \delta u + \frac{\partial g}{\partial v}(u,v) \cdot \delta v
\end{align*}
We call these defining equations \textit{overloading rules}.

\newcommand{\comout}[1]{}
\comout{

\newpage
\section{Heap}

Automatic differentiation aims at computing $\nabla f(\bx)\t$ and/or

We are concerned with the evaluation of first and second derivatives of a sufficiently smooth function
\begin{align*}
	f \ : \ \R^k \rightarrow \R\,,
\end{align*}
where $k \in \N$. We assume that $f$ is evaluated for an input $\bx\equiv (x_1,...,x_k)$ and yields the output $y$, which is $y=f(\bx)$. We write $\comp = \sqrt{-1}$ and $\be_j$ for the $i$th Cartesian vector.

We often use computational environments such as Matlab, where we want to evaluate derivatives of complex functions, potentially consisting of several thousands of lines of code.

\paragraph{Finite differences}
The easiest --- but also most error-prone --- way to compute derivatives is the use of finite differences. In this approach the value of $y$ is compared for calls of $f$ with a small perturbation in $\bx$. We find
\begin{align*}
	f(\bx + h \cdot \be_j) = f(\bx) + h \cdot \frac{\partial f}{\partial x_j}(\bx) + \cO(h^2)
\end{align*}
from which $\frac{\partial f}{\partial x_j}(\bx)$ can be estimated by the difference of $f(\bx + h \cdot \be_j)$ to $f(\bx)$. The problem with this approach is that rounding errors affect the values of $f(\bx + h \cdot \be_j)$, which leads to large relative rounding errors in the finite difference, cf. \cite{Squire:1998:UCV:278129.278138,Fornberg:1981:NDA:355972.355979}.

\paragraph{Complex step}
An easy and numerically reliable way to obtain first derivatives is to use complex perturbations. The idea is to perturb the input $\bx$ by a complex direction $\bv = h \cdot \comp \cdot \be_j$, where again $h>0$ is small. Under suitable conditions \cite{Martins:2003:CDA:838250.838251} the output $y = f(\bx+\bv)$ has a complex part that satisfies
\begin{align*}
\text{imag}(y) = \frac{\partial{}f}{\partial{}x_j}(\bx)\cdot h + \cO(h^3)\,.
\end{align*}
Iterating over all indices $j=1,...,k$ and choosing $h$ sufficiently small, we obtain accurate values for all first derivatives of $f$ in $\bx$. The rounding errors from finite differences do not appear because the scales of the (large) function value $f(\bx)$ and the (small) output perturbation do not interfere, since one is real-valued and the other is purely imaginary. For further details consult \cite{Martins:2003:CDA:838250.838251,Squire:1998:UCV:278129.278138}.

\paragraph{Dual numbers}
Instead of complex numbers one can implement dual numbers \cite{GriewankAutoDiff}, that consist of a \textit{primal number} $\bx$ which is augmented by a \textit{first order perturbation} $\delta\bx$. We denote the set of dual numbers with $\D$. We use the following notation for dual numbers: We write $\hat{y} = \langle y,\delta y\rangle \in \D$ and $\hbx = \langle \bx,\delta\bx\rangle \in \D^k$, where $y,\delta y \in \R$ and $\bx,\delta\bx \in \R^k$.

The idea is now that $f$ can be generalized such that it maps elements from $\D^k$ to $\D$ in the following way:
\begin{align*}
	\hat{y} &= f(\hbx) \\
	\langle y,\delta y\rangle &= f\big(\langle\bx,\delta\bx\rangle\big)
\end{align*}
There is a clear rule how $f$ has to be modified such that the following holds:
\begin{subequations}
\begin{align}
	y &= f(\bx)\\
	\delta y &= Jf(\bx)\cdot\delta\bx\,,
\end{align}\label{eqn:DualNumber}
\end{subequations}
where $Jf = (\partial{}f/\partial{}x_1,...,\partial{}f/\partial{}x_k)\in \R^{1 \times k}$ is the Jacobian of $f$.
\largeparbreak

Below we name this rule and illustrate it with an example. 
\begin{thm}[Recursive rule of dual numbers]
	All functions called within $f$ (no matter if they are subroutines, functions, or operators) must be redefined such that they satisfy the equations \eqref{eqn:DualNumber}. Then $f$ satisfies the equations \eqref{eqn:DualNumber}.
	\newline
	\underline{Proof:} A more general result is proven below.
\end{thm}

We give an example for this rule. Consider the following function for $f$.
\begin{algorithmic}
\Procedure{$f$}{$x_1,x_2$}
\State $w := x_1 + x_2$
\State $z := -x_2$
\State $y := w \cdot z$
\State \Return $y$
\EndProcedure
\end{algorithmic}
Rewriting $f$ as follows
\begin{algorithmic}
\Procedure{$f$}{$x_1,x_2$}
\State $w := \phi_{\text{add}}(x_1,x_2)$
\State $z := \phi_{\text{uni-minus}}(x_2)$
\State $y := \phi_{\text{times}}(w,z)$
\State \Return $y$
\EndProcedure
\end{algorithmic}
with functions
\begin{align*}
\phi_{\text{add}} \ &: \ \R^2 \rightarrow \R\\
\phi_{\text{uni-minus}} \ &: \ \R \rightarrow \R\\
\phi_{\text{times}} \ &: \ \R^2 \rightarrow \R\,,
\end{align*}
we find that the code for $f$ will be suitable for computations with dual numbers once after the functions $\phi_{\text{add}}$, $\phi_{\text{uni-minus}}$, $\phi_{\text{times}}$ have been generalized for dual numbers. Applying the rule related to \eqref{eqn:DualNumber}, we find the following generalizations:
\begin{align*}
	\phi_{\text{add}}\Big(\,\langle x_1,\delta x_1\rangle,\langle x_2,\delta x_2\rangle \,\Big)&=\langle x_1+x_2,\delta x_1 + \delta x_2\rangle\\
	\phi_{\text{uni-minus}}\Big(\,\langle x_1,\delta x_1\rangle\,\Big)&=\langle-x_1,-\delta x_1\rangle\\
	\phi_{\text{times}}\Big(\,\langle x_1,\delta x_1\rangle,\langle x_2,\delta x_2\rangle \,\Big)&=\langle x_1 \cdot x_2,x_1\cdot\delta x_2 + x_2\cdot\delta x_1\rangle
\end{align*}
Working through our example code of $f$ with these definitions, we find:
\begin{algorithmic}
	\Procedure{$f$}{\,$\langle x_1,\delta x_1\rangle,\langle x_2,\delta x_2\rangle$\,}
	\State $w := x_1 + x_2$,\quad $\delta w := \delta x_1+\delta x_2$
	\State $z := -x_2$,\quad $\delta z := - \delta x_2$
	\State $y := w \cdot z$,\quad $\delta y := w \cdot \delta z + z \cdot \delta w$
	\State \Return $\hat{y}\equiv\langle y,\delta y\rangle$
	\EndProcedure
\end{algorithmic}
Inserting all expressions yields
\begin{align*}
	\hat{y} = \big\langle\, (x_1+x_2) \cdot (-x_2) \,,\, -x_2 \cdot \delta x_1 - 2 \cdot x_2 \cdot \delta x_2 \,\big\rangle,
\end{align*}
which in turn satisfies the conditions \eqref{eqn:DualNumber}, just as Theorem~1 says.

}

\section{Hyper-dual numbers}
In a nested approach one could use dual numbers whose first order perturbations are dual numbers themselves. This could help for computing second-order derivatives of $f$. However, it makes the code difficult to read. To keep things simple, it seems more desirable to extend the mathematical framework of dual numbers.

Hyper-dual numbers \cite{Fike_HD} or hyper-dual vectors are an extension of dual numbers. Their purpose is to enable a more direct computation of second order derivatives.

Hyper-dual numbers consist of four numbers. We write $\hat{\hat{\bx}} = \langle x,\delta x_1, \delta x_2, \delta\delta x\rangle$ for a hyper-dual number, where $x,\delta x_1, \delta x_2, \delta\delta x \in \C$. We write $\H$ to denote the set of hyper-dual numbers. An $n$-dimensional vector of hyper-dual numbers we write as $\hat{\hat{x}} = \langle \bx,\delta \bx_1, \delta \bx_2, \delta\delta \bx\rangle \in \H^n$, where $\bx,\delta\bx_1,\delta\bx_2,\delta\delta\bx \in \C^n$.

The overloading rules for hyper-dual vectors are as follows:
\begin{defn}[Overloading rule for hyper-dual numbers]\label{def:HyperdualRule}
Let $g : \C^n \rightarrow \C$ be a basic function. Then the following equations define the overloaded function $g : \H^n \rightarrow \H$ for a call $\hat{\hat{y}} := g(\hat{\hbx})$.
\begin{subequations}
	\begin{align}
	y 				&:= g(\bx)\\
	\delta y_1 		&:= \nabla g(\bx)\t\cdot\delta\bx_1\\
	\delta y_2 		&:= \nabla g(\bx)\t\cdot\delta\bx_2\\
	\delta\delta y 	&:= \delta\bx_1\t\cdot\nabla^2g(\bx)\cdot\delta\bx_2 + \nabla g(\bx)\t\cdot\delta\delta\bx\,,
	\end{align}\label{def:eqn:Hyper-dualNumber}
\end{subequations}
\end{defn}
\noindent
This definition has been used in the recent books of \cite{Naumann:Art,GriewankAutoDiff}, but not in the original paper of Fike and Alonso \cite{Fike_HD} from 2011, where hyper-dual numbers are first introduced.

\paragraph{Using hyper-dual numbers}
Let us assume that we want to evaluate the Hessian of \texttt{foo} in a point $x \in \C$. To this end, we call \texttt{foo} with a hyper-dual number $\hat{\hat{x}} = \langle x, \delta x_1,\delta x_2,\delta\delta x\rangle$. In advance we have implemented the basic functions / operators $\sqrt{}$, $+$, $/$ for hyper-dual numbers according to Definition~\ref{def:HyperdualRule}\,. Due to these overloadings, the call of \texttt{foo} yields the following computations:
\begin{algorithmic}
	\Procedure{}{}{\texttt{foo}}({$\langle x,\delta x_1,\delta x_2,\delta\delta x\rangle$})
	\State $\langle u,\delta u_1,\delta u_2, \delta\delta u \rangle := \langle \sqrt{x} , 0.5 \cdot x^{-0.5} \cdot \delta x_1,0.5 \cdot x^{-0.5} \cdot \delta x_2,$
	\State $\phantom{\langle u,\delta u_1,\delta u_2, \delta\delta u \rangle := \langle }$ $\delta x_1 \cdot 0.25 \cdot x^{-1.5} \cdot \delta x_2 + 0.5 \cdot x^{-0.5} \cdot \delta\delta x \rangle$\\
	\State $\langle v, \delta v_1,\delta v_2,\delta\delta v\rangle := \langle x+u,\delta x_1 + \delta u_1, \delta x_2 + \delta u_2 , \delta \delta u + \delta \delta x \rangle$\\
	\State $\langle y,\delta y_1,\delta y_2,\delta\delta y \rangle := \langle v/u , (v \cdot \delta u_1 - u \cdot \delta v_1)/u^2, (v \cdot \delta u_2 - u \cdot \delta v_2)/u^2,$
	\State $\phantom{\langle y,\delta y_1,\delta y_2,\delta\delta y \rangle := \langle}$ $\begin{pmatrix}
	\delta v_1\\
	\delta u_1
	\end{pmatrix}\t \cdot \begin{bmatrix}
	0 & -\frac{1}{u^2} \\
	-\frac{1}{u^2} & \frac{2\cdot v}{u^3}
	\end{bmatrix}\cdot \begin{pmatrix}
	\delta v_2\\
	\delta u_2
	\end{pmatrix} + \begin{bmatrix}
	\frac{1}{u} & \frac{-v}{u^2}
	\end{bmatrix} \cdot \begin{pmatrix}
	\delta \delta v\\
	\delta \delta u
	\end{pmatrix}\rangle$
	\State \Return $\langle y , \delta y_1,\delta y_2,\delta\delta y \rangle$
	\EndProcedure
\end{algorithmic}
Inserting all the expressions into $y,\delta y_1,\delta y_2,\delta\delta y$, we find
\begin{align*}
	y &= \texttt{foo}(x)\\
	\delta y_1 &= \nabla \texttt{foo}(x)\t \cdot \delta x_1\\
	\delta y_2 &= \nabla \texttt{foo}(x)\t \cdot \delta x_2\\
	\delta\delta y &= \delta x_1\t \cdot \nabla^2 \texttt{foo}(x) \delta x_2 + \nabla \texttt{foo}(x)\t\cdot \delta\delta x\,.
\end{align*}
This means we can easily determine the Hessian of \texttt{foo} by making the call with $\delta x_1 = 1$, $\delta x_2 = 1$, $\delta\delta x = 0$ and reading $\delta\delta y$.

\paragraph{Proof of correctness of derivatives from hyper-dual numbers}
It strikes the eye that the equations for $\hat{\hat{y}}$, the result of the hyper-dual call of \texttt{foo}, do perfectly match with the defining equations \eqref{def:eqn:Hyper-dualNumber}. The following theorem proves that this is always the case.

\begin{thm}[Recursion of hyper-dual numbers]
Consider an implementation \texttt{f} of a function $f$. Let \texttt{f} interiorly call other functions, sub-functions, and/or operators, that satisfy the equations \eqref{def:eqn:Hyper-dualNumber}.
Then also \texttt{f} satisfies the equations \eqref{def:eqn:Hyper-dualNumber}.
\end{thm}
\noindent
\underline{Proof:} Consider the following functions.
\begin{align*}
f \ :& \ \H^k \rightarrow \H\\
g_j \ :& \ \H^n \rightarrow \H\quad\quad j=1,...,k\\
\psi:= f \circ \vec{g} \ :& \ \H^n \rightarrow \H
\end{align*}
It is sufficient to consider the following nested call:
\begin{align*}
\hat{\hat{z}} &= f\Big(\,g_1(\hat{\hbx}),...,g_k(\hat{\hbx})\Big)
\end{align*}
This is so because a nested call does cover all rules of differential calculus, such as chain-rule, product-rule and fraction-rule. To give a more practical explanation, we can interpret $g_1,...,g_k$ as functions that are called within the code of the implementation of $f$. The theorem makes the requirement that $g_1,...,g_k$ satisfy \eqref{def:eqn:Hyper-dualNumber}. What we now need to show is that $f \circ \vec g$ satisfies \eqref{def:eqn:Hyper-dualNumber}.

The function $\vec{g}$ is vectorial, consisting of components $g_j$, $j=1,...,k$. $\psi$ is the concatenation of $f$ and $\vec{g}$. For $\psi$ we can insert into \eqref{def:eqn:Hyper-dualNumber}. For $f$ and $g_j$ the equation \eqref{def:eqn:Hyper-dualNumber} instead works as a definition.

Our proof works as follows. Using \eqref{def:eqn:Hyper-dualNumber} for $f$ and $g_1,...,g_k$, we show that $\hat{\hat{z}}:=f(\vec{g}(\hat{\hbx}))$ is identical to the output of the call $\psi(\hat{\hbx})$.

We start by computing $\hat{\hat{\by}} := \vec{g}(\hat{\hbx})$. According to \eqref{def:eqn:Hyper-dualNumber}, we obtain:
\begin{align*}
\hat{\hat{\by}} =& \Big\langle\, \begin{pmatrix}
g_1(\bx)\\
\vdots\\
g_k(\bx)
\end{pmatrix}\,,\ \begin{pmatrix}
\nabla g_1(\bx)\t \cdot \delta\bx_1\\
\vdots\\
\nabla g_k(\bx)\t \cdot \delta\bx_1
\end{pmatrix}\,,\ \begin{pmatrix}
\nabla g_1(\bx)\t \cdot \delta\bx_2\\
\vdots\\
\nabla g_k(\bx)\t \cdot \delta\bx_2
\end{pmatrix}\,,\ \\
&\quad\quad\begin{pmatrix}
\delta\bx_1\t\cdot\nabla^2 g_1(\bx)\cdot\delta\bx_2+\nabla g_1(\bx)\t \cdot \delta\delta\bx\\
\vdots\\
\delta\bx_1\t\cdot\nabla^2 g_k(\bx)\cdot\delta\bx_2+\nabla g_k(\bx)\t \cdot \delta\delta\bx_1
\end{pmatrix}
\Big\rangle\\
=& \Big\langle\,\vec{g}(\bx)\,,\ \nabla\vec{g}(\bx)\t\cdot\delta\bx_1\,,\ \nabla\vec{g}(\bx)\t\cdot\delta\bx_2\,,\delta\bx_1\t\cdot\nabla^2 \vec{g}(\bx)\cdot\delta\bx_2+\nabla\vec{g}(\bx)\t\cdot\delta\delta\bx \,\Big\rangle
\end{align*}
Below we use $\hat{\hat{\by}}$ as input for $f$. According to \eqref{def:eqn:Hyper-dualNumber}, we obtain in turn:
\begin{align*}
\hat{\hat{z}}=&\Big\langle\,f(\by)\,,\ \nabla f(\by)\t\cdot\delta\by_1\,,\ \nabla f(\by)\t\cdot\delta\by_2\,,\ \delta\by_1\t\cdot\nabla^2 f(\by)\cdot\delta\by_2 + \nabla f(\by)\t\cdot\delta\delta\by\,\Big\rangle\\
=&\Big\langle\, f(\by)\,,\ \nabla f(\by)\t\cdot \nabla\vec{g}(\bx)\t\cdot\delta\bx_1\,,\ \nabla f(\by)\t\cdot \nabla\vec{g}(\bx)\t\cdot\delta\bx_2\,,\ \\
&\quad\quad \delta\bx_1\t\cdot \nabla\vec{g}(\bx)\cdot\nabla^2 f(\by) \cdot \nabla\vec{g}(\bx)\t\cdot\delta\bx_2 + \nabla f(\by)\t\cdot \nabla\vec{g}(\bx)\t\cdot\delta\delta\bx \\
&\quad\quad+ \nabla f(\by)\t\cdot\Big(\delta\bx_1\t\cdot\nabla^2\vec{g}(\bx)\cdot\delta\bx_2\Big) \,\Big\rangle\,,
\end{align*}
where we use the notation
\begin{align*}
\delta\bx_1\t\cdot\nabla^2\vec{g}(\bx)\cdot\delta\bx_2 := \begin{pmatrix}
\delta\bx_1\t\cdot\nabla^2 g_1(\bx)\cdot\delta\bx_2\\
\vdots\\
\delta\bx_1\t\cdot\nabla^2 g_k(\bx)\cdot\delta\bx_2
\end{pmatrix}\,.
\end{align*}
From vector algebra and differential calculus we find the identity
\begin{align*}
\nabla f(\by)\t\cdot\Big(\delta\bx_1\t\cdot\nabla^2\vec{g}(\bx)\cdot\delta\bx_2\Big) &= \delta\bx_1\t \cdot \Big(\sum_{j=1}^k \frac{\partial f}{\partial x_j}\big(\vec{g}(\bx)\big)\cdot\nabla^2 g_j(\bx) \Big)\cdot \delta\bx_2\\
&= \delta\bx_1\t \cdot \nabla^2 \psi(\bx) \cdot \delta\bx_2\,.
\end{align*}
We can simplify the expression for $\hat{\hat{z}}$.
\begin{align*}
\hat{\hat{z}} = \Big\langle\, \psi(\bx)\,,\ \nabla f(\bx)\t\cdot\delta\bx_1\,,\ \nabla f(\bx)\t\cdot\delta\bx_2\,,\ 
\delta\bx_1\t \cdot\nabla^2 \psi(\bx)\cdot \delta\bx_2 + \nabla f(\bx)\t\cdot\delta\delta\bx  \,\Big\rangle
\end{align*}
Thus, when $f$ and $\vec{g}$ satisfy \eqref{def:eqn:Hyper-dualNumber} then also $\psi$ does satisfy \eqref{def:eqn:Hyper-dualNumber}.\hspace{1cm} q.e.d.

Coming back to our implementation \texttt{f} of $f$, we draw the following conclusion from the theorem. If the code \texttt{f} is called with a hyper-dual vector as input then the output will by hyper-dual and satisfy the following equations
\begin{subequations}
	\begin{align}
	y 				&:= f(\bx)\\
	\delta y_1 		&:= \nabla f(\bx)\t\cdot\delta\bx_1\\
	\delta y_2 		&:= \nabla f(\bx)\t\cdot\delta\bx_2\\
	\delta\delta y 	&:= \delta\bx_1\t\cdot\nabla^2f(\bx)\cdot\delta\bx_2 + \nabla f(\bx)\t\cdot\delta\delta\bx\,,
	\end{align}\label{eqn:Hyper-dualNumber-f}
\end{subequations}
subject to that all functions that are interiorly called in the code of \texttt{f} satisfy the rule in Definition~\ref{def:HyperdualRule}.

The goal of an implementation of a hyper-dual number is to make sure that all basic functions and operators, that could potentially be called in a function code \texttt{f}, are overloaded according to  Definition~\ref{def:HyperdualRule}. This makes sure that the above theorem always applies and that the hyper-dual output of \texttt{f} satisfies \eqref{eqn:Hyper-dualNumber-f}.

\section{Free implementation in Matlab}
We have implemented a hyperdual number in Matlab as a class. It can be downloaded from \mbox{\texttt{http://www.martinneuenhofen.de/HyperDual/HyperDual.html}}\,. The code provides functions for the construction and access of hyper-dual vector-valued objects. Further, a wide range of elementary functions and operators in Matlab (e.g. plus, times, as well as trigonometric and power functions) has been implemented.

Apart from the implementation of the hyper-dual number itself, our code offers two interfaces that are easy to use for the evaluation of the Jacobian and/or Hessian of a Matlab function \texttt{f}.

When the function $f$ has an input dimension $n>1$, multiple hyper-dual function calls of \texttt{f} are required to evaluate the full Jacobian and Hessian of $f$. Given an m-file with a function code, like for instance below,
\begin{verbatim}
function y = f(x) % k=2
y  = x(1) + x(2).^2 .* x(3) - x(1)./x(3) + x(2).^x(1);
% x(2) must be positive
end
\end{verbatim}
the following calls write the Jacobian and Hessian of \texttt{f} in the exemplary point $\bx = (1+\sqrt{-1},2.3,\pi)\t$ into \texttt{Dy} and \texttt{Hy}. 
\begin{verbatim}
x  = [ 1+sqrt(-1) ; 2.3 ; pi ];
Dy = HD_Jacobian_Call(@f,x);
Hy = HD_Hessian_Call(@f,x);
\end{verbatim}
These functions work as follows. They call \texttt{f} repeatedly with hyper-dual numbers of the form $\hat{\hbx} = \langle \bx , \be_j,\be_k,\bO\rangle$, where $\be_\ell$ is the $\ell$th Cartesian unit vector. Iterating for $j=1,...,n$, $k=1,...,j$, it follows
\begin{align*}
	\delta y_1 &= \frac{\partial f}{\partial x_j}(\bx)\\
	\delta\delta y &= \frac{\partial^2 f}{\partial x_j\,\partial x_k}(\bx)
\end{align*}
I.e. the values of the Jacobian and Hessian are readily available from the output values of the hyper-dual call of \texttt{f}. 

\paragraph{Computational cost}
For the evaluation of the whole Jacobian matrix there are $n$ calls of \texttt{f} required. More computationally prohibitive, the evaluation of the whole Hessian matrix requires $n\cdot(n-1)/2$ calls. This is disadvantageous compared to an adjoint differentiation approach, which can compute a Hessian matrix in $n$ function calls and a Jacobian in a single function call, cf. \cite{GriewankAutoDiff,Naumann:Art,Martins:2002:CAM}. However, it is yet a topic of active research to make adjoint differentiation algorithms as robust as dual numbers \cite[p. 2]{Fike_HD}.

\section{Concluding remarks}
We have introduced hyper-dual numbers as a simple and robust algorithmic concept for the computation of first and second order derivatives. We presented a theorem that proves the correctness of the values for the Jacobian and Hessian matrices as computed with hyper-dual numbers. We provided a simple implementation of hyper-dual numbers in Matlab that is freely available and easy to use.

\FloatBarrier

\bibliography{hyperdualBib}
\bibliographystyle{plain}

\end{document}